# GAMMA RAY BURST TRIGGERING SUPERNOVA EXPLOSION (AND OTHER EFFECTS ON NEIGHBOURING STARS)

C. Sivaram and Kenath Arun

Indian Institute of Astrophysics, Bangalore

**Abstract:** The initial burst of a gamma ray burst (GRB) is usually followed by a longer-lived afterglow emitted at longer wavelengths. The evidence for a physical connection between GRBs and core collapse supernovae (SN) has increased since the discovery of GRB afterglows [1, 2]. So far SN signatures have been found in only a few GRBs. Here we propose the possibility of a GRB triggering the collapse of a WR or RG star in a binary system producing a SN, and typical signatures. We also look at the effects of GRBs on MS and WD stars in the neighbourhood. The possibility of GRBs retarding star formation in an interstellar cloud is also discussed.

Gamma-ray bursts (GRBs) are flashes of gamma rays associated with extremely energetic explosions and can last from milliseconds to several minutes [3, 4]. It is generally believed that the long-duration gamma ray bursts are associated with the deaths of massive stars in a kind of supernova-like event such as a hypernova [5]. Here we propose that a GRB in a binary system with a progenitor of SN, such as a Wolf-Rayet star or massive red supergiant star could trigger the collapse of the WR or RG star producing a supernova explosion. [6]

The gamma-ray energy released in a typical GRB is of the order of $\sim 10^{52}$ ergs. The flux of gamma rays at a distance of about 50AU is:

$$f = \frac{E_{GRB}}{4\pi D^2} \approx 10^{21} \text{ ergs/cm}^2\text{/s} \qquad \text{... (1)}$$

And the corresponding energy density is given by:

$$\varepsilon = \frac{f}{4c} \approx 10^{10} \text{ ergs/cc} \qquad \text{... (2)}$$

The total gamma-ray energy falling on the RG (or WR) in a duration of several seconds will be far more (trillion times) than the bolometric radiation emitted by the star. Hence this flux of gamma rays can induce the collapse of the RG star to produce a SN. On the RG the energy falling $\sim 10^{50}$ ergs (on a WR star it is $\sim 10^{45}$ ergs).

For a red supergiant, like Betelgeuse, the gravitational pressure $\left(\approx \frac{GM^2}{R^4}\right)$~1dyne/cm$^2$. The isotropic gamma-ray flux from the GRB falling on the red supergiant would be $10^{10}$ times more, or if it is focused, would be $10^{13}$ times larger!

If the source of gamma rays is collimated into a beam at an opening angle of ~$5^0$, then the flux received by the companion star will be more by a factor of ~$10^3$ and the energy received will be focused in an area of ~$10^{28}$ cm$^2$.

The time taken for the collapse is given by:

$$\tau = \left(\frac{R^3}{GM}\right)^{1/2} \sim 100 \text{ days} \qquad \text{... (3)}$$

for RG and a few days in the case of WR star.
This could possibly explain the lag in observing the associated SN signature after the GRB. [7]

In case the companion star is another WR star of similar mass, metallicity and rotation, there is a possibility that it may also collapse into a black hole giving rise to another gamma ray burst. So a possible signature of such a scenario (i.e. a WR binary, where one of the stars collapses to cause a GRB) is two consecutive GRB's separated by perhaps several days, from the same region of the sky, which at a Gpc distance corresponds to an angular separation of $\left(\sim \frac{10^{15}}{10^{28}} \text{ rad}\right)$ ~0.02 microarcsec!

Of course gamma ray resolution is nowhere near this figure, but if there are optical afterglows associated with both explosions and these spread over $10^{15}$m, then they would be just within the current interferometric precision of about 20 microarcsec! So we should see two afterglows within 20 microarcsec separated by several weeks! [8]

Again if the companion is a red SG, then the resulting core collapse SN would again be spatially separated by a picoarcsec with a time lag of a few months! Perhaps future space telescopes of the ATLAS type or the OWL on the ground could be sensitive to these phenomena.

If solar like main sequence stars, or WD's are within few light years from such a GRB (occurring in dense star forming regions) then the gamma ray flux on a MS star is:

$$f = \frac{E_{GRB}}{4\pi D^2} \approx 10^{14} \text{ ergs/cm}^2\text{/s} \qquad \text{... (4)}$$

(Total gamma ray irradiation would be ~$10^{37}$ergs ~$10^3$ times its bolometric luminosity!)

A white dwarf close to the Chandrasekhar limit at a distance of ~1 light-year from the GRB would experience a gamma ray irradiation of ~$10^{33}$ ergs, heating it to a temperature:

$$T = \left(\frac{f}{\sigma}\right)^{1/4} > 10^5 K \qquad \text{... (5)}$$

perhaps giving rise to a nova like outburst. At that distance this may not generate a collapse of the WD.

A main sequence star at a distance of a few light-years from the GRB would be irradiated with a gamma ray flux of

$$f = \frac{E_{GRB}}{4\pi D^2} \approx 10^{13} \text{ ergs/cm}^2\text{/s} \qquad \text{... (6)}$$

and could exhibit high energy flares, of several thousand times their bolometric luminosity for several minutes, with a corresponding increase in temperature

$$T = \left(\frac{f}{\sigma}\right)^{1/4} > 10^4 K \qquad \text{... (7)}$$

Although GRB's could have drastic effects on neighbouring stellar objects, they could inhibit continuing star formation, as even an interstellar gas cloud hundred parsecs away could be heated to several thousand degrees and dissipate. For example, a cloud of $10^2$ solar mass, with average density of $10^2$ atoms/cc would be in virial equilibrium only at T ~20K and dust grains could be heated and vaporised by the gamma ray flux. [9]